\documentclass[
 amsmath,amssymb,
 aps,
 amsmath,amssymb,
 superscriptaddress,
 reprint,%
]{revtex4-2}

\usepackage[dvipdfmx]{graphicx}
\usepackage{dcolumn}
\usepackage{bm}
\usepackage{here}
\usepackage{color}

\begin{document}
\newcommand{\noter}[1]{{\color{red}{#1}}}

\preprint{APS/123-QED}

\title[Sample title]{Enhanced collective vibrations in granular materials}

\author{Shihori Koyama}
\email{koyama-shihori146@g.ecc.u-tokyo.ac.jp}
\affiliation{Graduate School of Arts and Sciences, The University of Tokyo, Tokyo 153-8902, Japan}
\author{Norihiro Oyama}
\email{Norihiro.Oyama.vb@mosk.tytlabs.co.jp}
\affiliation{Toyota Central R\&D Labs., Inc., Nagakute, Aichi 480-1192, Japan}
\author{Hideyuki Mizuno}
\email{hideyuki.mizuno@phys.c.u-tokyo.ac.jp}
\affiliation{Graduate School of Arts and Sciences, The University of Tokyo, Tokyo 153-8902, Japan}
\author{Atsushi Ikeda}
\email{atsushi.ikeda@phys.c.u-tokyo.ac.jp}
\affiliation{Graduate School of Arts and Sciences, The University of Tokyo, Tokyo 153-8902, Japan}
\affiliation{Research Center for Complex Systems Biology, Universal Biology Institute,
The University of Tokyo, Tokyo 153-8902, Japan}


\begin{abstract}
Granular materials are defined as collections of macroscopic dissipative particles. 
Although these systems are ubiquitous in our lives, the nature and the causes of their non-trivial collective dynamics still remain elusive and have attracted significant interest in non-equilibrium physics.
Here, we focus on the vibrational dynamics of granular materials. 
While the vibrational dynamics of random packings have been examined concerning the jamming transition, previous research has overlooked the role of contact dissipations.
We conducted numerical and analytical investigations into the vibrational dynamics of random packings influenced by the normal dissipative force, which is the simplest model for contact dissipations. 
Our findings reveal that the kinetic energy per mode diverges in the low-frequency range, following the scaling law $\mathcal{K}_l \propto \omega^{-2}_l$ with the frequency $\omega_l$, indicating that low-frequency modes experience strong excitation and that the equipartition of energy is violated. 
Additionally, the spatial structure factor of the velocity field displays the scaling law $S_v(q) \propto q^{-2}$ with the wavenumber $q$, which signifies that the velocity field has an infinitely long range.
We demonstrate that these phenomena arise from the effects of weaker damping on softer modes, where the particle displacements parallel to the contacts are minimal in the low-frequency modes, rendering normal dissipation ineffective at dampening these modes.
\end{abstract}

\maketitle

\section{Introduction}
Granular materials, such as sand and flour, are defined as collections of macroscopic dissipative particles.
While these systems are ubiquitous in our lives, their dynamics are highly complex, and many questions remain unanswered from a physics perspective~\cite{JN1992,deGennes1999}.
Understanding their dynamics is crucial for engineering and other natural sciences as they provide guidelines for handling granular materials in industrial processes and also offer insights into various natural phenomena, such as snow avalanches and the formation of planetary rings~\cite{BGT1985,JNB1996,DD1999}.
A key feature of the dynamics of granular materials is the emergence of collective behavior~\cite{AT2006}.
In dilute systems, granular particles in a free cooling state tend to form clusters, and the spatial heterogeneity results in non-trivial energy decay~\cite{GZ1993,PJD2014}.
In dense systems, granular particles in sheared or vibrated states exhibit vortex structures in their velocity fields, which have been an important research focus~\cite{GHS1992,CSG2004,JFP2006,ARO2012,FOH+2015,KN2016,Windows2016,KR2017,KT2018}.
Because granular materials are dissipative, these macroscopic dynamics emerge as non-equilibrium phenomena that respond to external manipulations of the systems~\cite{AS1973}.

The physics of the jamming transition provides a useful framework for understanding the complex dynamics of dense granular materials.
Random packings of athermal, frictionless particles serve as the central model in this area of research~
\cite{OSL+2003}.
This model achieves mechanical stability at a specific density, known as the jamming transition~\cite{PK1986,BW1990,OSL+2003,MSL+2007,vanHecke2009,CCS+2009,BRA+2010,BC2018}.
The jamming transition represents a form of critical phenomenon, where various physical properties exhibit power laws in relation to the proximity to the critical density~\cite{OLL+2002,OSL+2003,vanHecke2009}.
Among these, the vibrational dynamics of random packings have been extensively studied.
The vibrational modes of the packings are derived from the diagonalization of the dynamical matrix, which depends solely on the mass, inter-particle interactions, and the configurations of the constituent particles.
In crystals, these vibrational modes manifest as plane waves, and the vibrational density of states adhere to the notable Debye law $D(\omega) \propto \omega^{d-1}$, where $d$ denotes the spatial dimension.
In contrast, many vibrational modes of the random packings are spatially disordered, contributing to an excess density of states relative to the Debye law.
$D(\omega)$ exhibits a plateau down to a characteristic frequency that reveals critical power-law behavior near the jamming transition.
Furthermore, the spatially localized vibrational modes, referred to as quasi-localized modes, are known to emerge in the lowest frequency regime~\cite{MSI2017,KBL2018,SMI2018}.
Importantly, the linear rheological properties of the packings are dictated by these vibrational modes~\cite{LM2006}, highlighting the significance of understanding their vibrational properties. 

In parallel to these numerical and theoretical works on vibrational properties near jamming, the vibrational dynamics of dense granular materials have been studied experimentally.
One method involves recording the covariance matrix of grain positions and diagonalizing it to obtain the vibrational modes.
This analysis captures collective vibrations in the system~\cite{BDBB2010,HBD2012}.
Another method records the time correlation function of grain velocities and transforms it into the frequency domain.
Through this approach, an increase in low-frequency vibrations near jamming was observed~\cite{OD2013,BWO2024}.
However, the connection between these experimental observations on granular materials and the above-mentioned understanding of vibrational modes of random packings remains unclear~\cite{SvHE+2007,SSL2019}.
This arises because conventional vibrational mode analysis has neglected dissipation, which is present in experimental granular systems.
Dissipation can alter the amplitude of vibrational mode excitations or even the vibrational modes themselves.
To fully understand the vibrations of granular materials, it is essential to explicitly incorporate dissipation into vibrational mode analysis~\cite{HBD2012,CGP+2019,ISO+2023}. 

In this work, we study the vibrational dynamics of a model of granular materials. We examine random packings of particles driven by external random forces as the simplest setup.
In particular, we demonstrate the non-trivial roles played by dissipation in vibrational dynamics by comparing two distinct systems that share the same setups but the precise implementation of dissipative forces.
The first system is the simplest model for granular materials, in which the effects of inelastic collisions are incorporated via normal dissipative forces. In the second system, dissipation is implemented by the Stokes drag force.
Although this approach is well-suited to describing colloidal suspensions, we study it as a reference system.
We solve the linearized equations of motion for these systems both analytically and numerically.
Our findings indicate that the precise form of the dissipation significantly alters the vibrational dynamics.
Notably, low-frequency vibrations are markedly amplified and long-wave-length collective motions arise in the case of the normal dissipative forces, whereas the law of equipartition of energy is satisfied under the Stokes drag.

This paper is organized as follows.
We begin by introducing our models in Section~\ref{sec:model_methods}.
The simulation methods and physical observables are also presented in this section.
Next, in Sec.~\ref{sec:ke}, we analyze the excitations of vibrational modes using the analytical solutions of the linearized equations of motion.
We demonstrate that the low-frequency modes are strongly excited in the case of normal dissipative force.
In Sec.~\ref{sec:sp_structure}, we discuss the spatial properties of the vibrations.
In particular, we show that collective motions arise due to the strong excitation of low-frequency modes.
\section{Model and Methods\label{sec:model_methods}} 
\subsection{Model\label{subsec:model}}

In this section, we describe our granular models. 
We consider a three-dimensional monodisperse system consisting of $N$ particles with equal masses $m$. 
The system obeys the equation of motion:
\begin{align}
\label{eq:eom}
m \frac{d^2\bm{r}_i}{dt^2} = {\bm{F}}_i^C + {\bm{F}}_i^D + {\bm{F}}_i^R,
\end{align}
where ${\bm{r}}_i =(r_{ix}, r_{iy}, r_{iz})^\top$ is the position of particle $i$, and $\top$ denotes transpose operation.
The vectors ${\bm{F}}_i^C, {\bm{F}}_i^D, {\bm{F}}_i^R$ denote the conservative force, dissipative force, and random force acting on particle $i$, respectively.

In our model, the particles interact through the harmonic potential~\cite{Durian1995}:
\begin{align}
    \Phi = \sum\limits_{ij}\frac{1}{2}k(\sigma - r_{ij})^2 \Theta(\sigma - r_{ij}),
\end{align}
where $k$ is the spring constant, $\sigma$ is the particle diameter, $r_{ij}=|{\bm{r}}_{ij}|=|{\bm{r}}_i - {\bm{r}}_j|$ is the distance between particles $i$ and $j$, and $\Theta(x)$ is the Heaviside step function. 
Therefore, particles $i$ and $j$ interact only when they overlap $r_{ij} < \sigma$. 
Then, the conservative force acting on particle $i$ is given by
\begin{align}
    {\bm{F}}_i^C = - \frac{\partial \Phi}{\partial {\bm{r}}_i}.
\end{align}

For the dissipative force, we consider two different models. 
One is the normal dissipative force model:
\begin{align}
\label{eq:normal}
{\bm{F}}_i^D = -\eta_n \sum\limits_{j \in \partial_i} {\bm{n}}_{ij}{\bm{n}}_{ij}^\top ({\bm{v}}_i - {\bm{v}}_j). 
\end{align}
$\partial_i$ denotes the set of particles in contact with particle $i$, and $\eta_n$ is the coefficient of the normal dissipative force. 
${\bm{v}}_i$ is the velocity of particle $i$ and ${\bm{n}}_{ij}={\bm{r}}_{ij}/r_{ij}$ is the unit vector along the contact between particles $i$ and $j$.
Therefore, this dissipative force is proportional to the relative velocity parallel to the contact. 
This is the simplest model of inelastic collisions of granular materials. 
The other dissipative force model in our research is the Stokes drag model:
\begin{align}
\label{eq:stokes}
    {\bm{F}}_i^D = - \eta_s {\bm{v}}_i,
\end{align}
where $\eta_s$ represents the coefficients of the Stokes drag. 
Although this is the simplest model of the dissipation of colloidal particles immersed in solvent, we study this model as well, for comparison.  

For the external force, we consider the random white noise satisfying the following property:
\begin{align}
\langle {\bm{F}}_i^R(t){\bm{F}}_j^{R \top}(s) \rangle = 2B\delta_{ij}\delta(t-s) \bm{I}_3,
\label{eq:white}
\end{align}
where $B$ is a parameter that determines the strength of the white noise, $\delta_{ij}$ is the Kronecker's delta, $\delta(t)$ is the Dirac's delta function, and $\bm{I}_3$ is the $3\times 3$ identity matrix. $\langle\cdot\rangle$ represents the average over noise. 
Note that this work focuses on the random force for its simplicity. 
In real granular materials, there is a wide variety of energy injections, such as vertical vibrations, shear deformations, and air fluidization. 
Some of them are characterized by the specific frequency and the resulting dynamics can depend on it. 
To avoid these complications, we focus on the random force as it has the simplest flat frequency characteristics.

\subsection{Linearization of the equation of motion\label{subsec:linearizaion}}

We consider the model at the volume fraction of $\phi \simeq 0.74$ which is higher than the jamming density. 
Our analyses focus on such a high density system to reveal general features of the vibrations of jammed granular materials without paying too much attention to the criticality of the jamming transition.
We focus on a mechanically stable configuration, or in other words, an inherent structure of the model. 
The position of particle $i$ in the inherent structure is denoted by ${\bm{R}}_i=(R_{ix}, R_{iy}, R_{iz})^\top$. 
Hereafter, we introduce a compact notation of a $3N$-dimensional vector like ${\bm{a}}=(\bm{a}_1^\top, \bm{a}_2^\top,\cdots, \bm{a}_N^\top)^\top$. 
For example, the positions of particles in the inherent structure are denoted by ${\bm{R}}=(\bm{R}_1^\top, \bm{R}_2^\top,\cdots, \bm{R}_N^\top)^\top$. 

When the random force is sufficiently weak, the model behaves as a solid and the particles move only around their positions in the inherent structure. In this situation, the prestress (or pressure) keeps the particles in contact with each other and the interparticle contacts never be broken by random forces even if the dissipative forces are absent. Hence, we linearize the equation of motion Eq.~\eqref{eq:eom} in terms of the displacements of the particles, ${\bm{u}}\equiv{\bm{r}}-{\bm{R}}$, to obtain: 
\begin{align}
\label{eq:eom_linearized}
 m  \frac{d^2\bm{u}}{dt^2} + \Gamma\frac{d\bm{u}}{dt} + \mathcal{H}{\bm{u}} = {\bm{F}}^R,
\end{align}
where $\Gamma$ is the $3N\times3N$ damping matrix and $\mathcal{H}$ is the $3N\times3N$ hessian matrix of the potential, which will be given below. 

From the definition of the dissipative forces Eqs.~\eqref{eq:normal} and \eqref{eq:stokes}, the damping matrix $\Gamma$ is composed of the $3\times3$ matrix elements $\Gamma_{ij}$ corresponding to the pairs of particles $i$ and $j$.
More specifically, the damping matrix for the normal dissipative force model Eq.~\eqref{eq:normal} is given by 
\begin{align}
\label{eq:gamma_normal}
\Gamma_{ij}= \begin{cases} \eta_n \sum\limits_{l \in\partial_i} {\bm{n}}_{il}{\bm{n}}_{il}^\top \,\,\,(i=j),\\
- \eta_n {\bm{n}}_{ij}{\bm{n}}_{ij}^\top \,\,\,\,\,(j\in \partial_i), \\
{\bm{O}}_3\,\,\,\,\,(\mathrm{others}), 
\end{cases}
\end{align}
where ${\bm{O}}_3$ is the $3\times 3$ zero matrix.
The damping matrix for the Stokes drag model Eq.~\eqref{eq:stokes} is given by
\begin{align}
\label{eq:gamma_stokes}
\Gamma_{ij}= \begin{cases} \eta_s {\bm{I}}_3\,\,\,(i=j),\\
{\bm{O}}_3\,\,\,\,\,(i\ne j).
\end{cases}
\end{align}
%

Similarly, the hessian matrix $\mathcal{H}$ is composed of $3\times 3$ matrix elements $\mathcal{H}_{ij} = \frac{\partial^2\Phi}{\partial{\bm{r}}_i \partial{\bm{r}}_j^\top}|_{{\bm{r}}={\bm{R}}}$ corresponding to the pairs of particles $i$ and $j$. 
In general, the element $\mathcal{H}_{ij}$ depend on both the first derivative of the potential $\frac{d\phi_{ij}}{dr_{ij}}$ and the second derivative $\frac{d^2\phi_{ij}}{dr_{ij}^2} = k$. 
Here, we employ a simple approximation $\frac{d\phi_{ij}}{dr_{ij}}=0$, which is called the ``unstressed'' approximation. 
In this approximation, all the contacts between particles are replaced by the relaxed springs with the spring constant $k$. 
Then the element of the hessian matrix is given by: 
\begin{align}
\label{eq:unstressed_dm}
\mathcal{H}_{ij}=\begin{cases}
k \sum\limits_{l \in\partial_i} {\bm{n}}_{il}{\bm{n}}_{il}^\top \,\,\,(i=j),\\
- k {\bm{n}}_{ij}{\bm{n}}_{ij}^\top \,\,\,\,\,(j\in \partial_i),\\
{\bm{O}}_3\,\,\,\,\,(\mathrm{others}). 
\end{cases}
\end{align}

In summary, the linearized equation of motion is controlled by the damping matrix $\Gamma$ and the hessian matrix $\mathcal{H}$. 
These matrices are fully determined by the inherent structure. 
In the following sections, we solve this equations of motion for the normal dissipative force model~\eqref{eq:gamma_normal} and Stokes drag model~\eqref{eq:gamma_stokes} to discuss the vibrations of granular materials.  

\subsection{Observables\label{subsec:observables}}

In this section, we introduce the observables to characterize the vibrations in our granular model. 
We first introduce the eigenequations of the hessian matrix: 
\begin{align}
\mathcal{H}{\bm{e}}_l = \lambda_l {\bm{e}}_l, 
\end{align}
where ${\bm{e}}_l$ is the $l$-th eigenvector and $\lambda_l$ the $l$-th eigenvalue~($l=1,2,\cdots,3N$). 
Physically, $\lambda_l$ describes the effective spring constant for the mode $\bm{e}_l$. 
Accordingly, we introduce the characteristic frequency of the $l$-th mode as 
\begin{align}
\omega_l = \sqrt{\frac{\lambda_l}{m}}. 
\end{align}
For crystals and structural glasses where the dissipation does not play important role, these eigenvectors and eigenvalues carry all the relevant information of the vibrational dynamics of the system. 
In particular, thermal agitation excites these modes with the same energy, which is known as the law of equipartition of energy. 
However, in our granular model, the non-trivial dissipation is present, which may break the law of equipartition. 

To quantify the excitations of the modes, we focus on the kinetic energy associated with each mode. 
To this end, we expand the displacement fields using the modes as 
\begin{align}
\bm{u} = \sum_{l=1}^{3N} C_l \bm{e}_l.
\end{align}
Here, $C_l = {\bm{e}}_l^\top {\bm{u}}$ is the displacement along the $l$-th mode, and also $\dot{C}_l = dC_l/dt$ is the velocity along the $l$-th mode. 
Then the kinetic energy of the $l$-th mode is given by 
\begin{align}
\label{eq:kinetic_energy}
\mathcal{K}_l = \frac{1}{2} m \langle \dot{C}_l^2 \rangle. 
\end{align}
We calculate $\mathcal{K}_l$ for our granular models to examine the excitation of each mode. 

Next, to quantify the spatial structure of the vibrational dynamics, we introduce a spatial correlation function of the velocities of particles~\cite{HKC+2020}: 
\begin{align}
\label{eq:Sv}
    S_v(q)=\langle \bm{v}_{\bm{q}} \cdot \bm{v}_{-\bm{q}} \rangle. 
\end{align}
Here, $\bm{v}_{\bm{q}} = \frac{1}{\sqrt{N}}\sum \limits_{i=1}^N \bm{v}_i \exp(-i\bm{q}\cdot \bm{R}_i)$ is the Fourier transform of the velocity field of the system, $\bm{q}$ is the wave vector, and $\bm{v}_i = d \bm{u}_i /dt$ is the velocity of the particle $i$. If $S_v(q) \to \infty$ in the limit of large scale $q\to 0$, it means that the spatial structure expands whole the system.

Finally, to quantify the alignment of the velocities of neighboring particles, we introduce the following parameter: 
\begin{align}
\label{eq:op_v}
    {\Psi}_{i} = \frac{1}{N_i}\sum\limits_{j \in \partial_i}\frac{{\bm{v}}_i\cdot {\bm{v}}_j}{|{\bm{v}}_i||{\bm{v}}_j|},
\end{align}
where $N_i$ is the number of particles in the set $\partial_i$. 
This parameter quantifies the tendency of the alignment of velocities of neighboring particles. 
If $\Psi_{i}=1$, particle $i$ moves in the same direction as its neighboring particles.
If $\Psi_{i}=0$, particle $i$ and its neighboring particles move in completely random directions.

\subsection{Simulation setup\label{subsec:simlation}}
We first obtained the inherent structure of the model in the same way as Ref.~\cite{MSI2017}. 
The rattler particles which have less than $4$ contacting particles were removed iteratively. 
Starting from the inherent structures, we performed the molecular dynamics simulations of the linearized equation of motion Eq.~\eqref{eq:eom_linearized}. We performed sufficiently long simulations to obtain the steady state of the system, and then performed product runs where we calculated the physical quantities defined in Sec.~\ref{subsec:observables}. 
To calculate the kinetic energies introduced by Eq.~\eqref{eq:kinetic_energy}, we diagonalized the hessian matrix $\mathcal{H}$ to obtain the eigenvectors and the eigenfrequencies.
We set the dissipative force parameter to $\eta_s=\eta_n=0.158\sqrt{mk}$ which corresponds to typical restitution coefficient of granular particles $\varepsilon \simeq 0.7$~\cite{LK1964,SSC+2021}.
The random force parameter was set to $B=0.0001 \sigma^2\sqrt{mk^3}$ which justifies the linear approximation of the equation of motion. This corresponds to the situation in which the random forces are vanishingly weak but large enough for macroscopic motion to be clearly observed.

To integrate the equation of motion, we utilized the DPD-VV method~\cite{BVK+2000}, which is an extension of the Velocity-Verlet method to particle systems with dissipation. 
The time step of the simulation was set to $\Delta t = 0.005 \sqrt{m/k}$. 
The models with several different system sizes are considered: $1000 \leq N \leq 512000$. These system sizes enable us to observe the spatially heterogeneous vibrational dynamics. 
For each system size, we prepared 500 different steady-state configurations for $N \leq 125000$ and 250 configurations for $N=256000, 512000$. 
The results are all averages over these samples.
Note that we used the same inherent structures for both the Stokes drag model and the normal dissipative force model. 
\section{\label{sec:ke} Kinetic energy on vibrational mode}
In our models, the hessian matrix $\mathcal{H}$ and the damping matrix $\Gamma$ share the same matrix structure. 
This enables us to solve the equations of motion of our models. 
In the following, we calculate the kinetic energies using the analytical solutions of the equations of motion. 
\begin{figure}[tb]
  \centering
  \includegraphics[keepaspectratio, width=\linewidth]{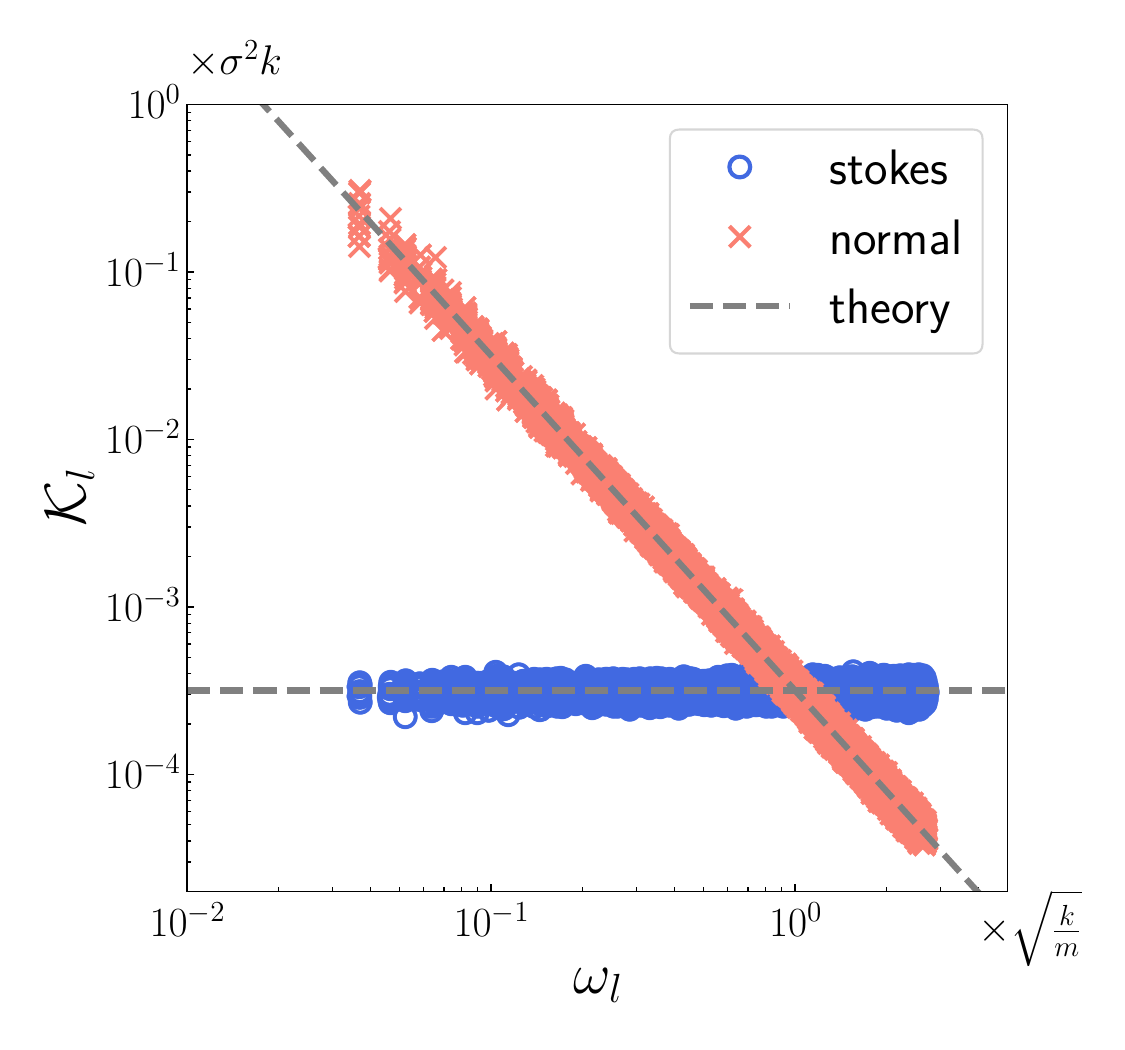}
  \caption{
The kinetic energy for each mode against the characteristic frequency. 
Symbols express the simulation results and lines represent the theoretical results Eqs.~\eqref{eq:ke_stokes} and \eqref{eq:ke_normal}. 
Blue symbols are for the Stokes drag model and red symbols are for the normal dissipative force model. 
The law of equipartition of energy is violated for the later case. } 
\label{fig:energy}
\end{figure}
\subsection{\label{subsec:stokes}Stokes drag model}

We first focus on the Stokes drag model Eq.~\eqref{eq:gamma_stokes}. 
Since $\Gamma \bm{e}_l = \eta_s \bm{e}_l$ in this case, we can expand the equation of motion into each mode $l$~($=1,2,\cdots,3N$) as:
\begin{align}
\label{eq:eom_mode}
m \ddot{C}_l + \eta_s \dot{C}_l + \lambda_l C_l = f_l, 
\end{align}
where $f_l$ is the random force for the $l$-th mode, defined by $\bm{F}^R = \sum_{l=1}^{3N} f_l \bm{e}_l$. 
This equation of motion can be solved by the variation of constants method (see Appendix A). 
Using this solution, we obtain the kinetic energy for each mode as: 
\begin{align}
\label{eq:ke_stokes}
\mathcal{K}_l 
= \frac{1}{2}m\langle \dot{C}_l^2 \rangle
= \frac{B}{2\eta_s}.
\end{align}
Therefore, all the modes are equally excited in the Stokes drag model. 
This is reasonable since this model can be seen as colloidal particles in a solvent, which are in the thermal equilibrium state.

To confirm this result, we performed MD simulations for the Stokes drag model (see Sec.~\ref{subsec:simlation} for details). 
The obtained kinetic energy $\mathcal{K}_l$ are shown against the characteristic frequency $\omega_l$ as the blue symbols in Fig.~\ref{fig:energy}. 
Clearly, the simulation result agrees with the theoretical result Eq.~\eqref{eq:ke_stokes} shown by the dashed line.

\subsection{\label{subsec:normal}Normal dissipative force model}
Next, we consider the normal dissipative force model Eq.~\eqref{eq:gamma_normal}. 
For this case, it is crucial to observe that the hessian matrix Eq.~\eqref{eq:unstressed_dm} and the damping matrix Eq.~\eqref{eq:gamma_normal} have the same matrix structure and they are proportional $\Gamma = \frac{\eta_n}{k} \mathcal{H}$. 
Therefore, the two matrices share the same eigenvectors and the following relation holds: 
\begin{align}
\label{eq:normaleigen}
\Gamma \bm{e}_l = \frac{\eta_n}{k} \mathcal{H} \bm{e}_l = \frac{\eta_n \lambda_l}{k} \bm{e}_l. 
\end{align}
This result means that the effective damping coefficient for the $l$-th mode is $\eta_n \lambda_l /k$. 
Since $\lambda_l$ is the effective spring constant for the $l$-th mode, Eq.~\eqref{eq:normaleigen} means that the damping becomes weaker as the mode is softer. 
This is natural due to the nature of the normal dissipative force. 
The hessian matrix Eq.~\eqref{eq:unstressed_dm} describes the energy cost that comes from the relative displacements parallel to the contacts. 
If $\lambda_l$ is small, such parallel displacements are small in the mode $l$. 
Since the normal dissipation damps the relative motions parallel to the contacts, the effective damping coefficient should be smaller when the parallel displacements are smaller.
This effect, that is weaker damping for softer mode, is described by Eq.~\eqref{eq:normaleigen}. 

Using Eq.~\eqref{eq:normaleigen}, the equation of motion can be decomposed into each mode as:
\begin{align}
\label{eq:eom_normal}
m \ddot{C}_l + \frac{\eta_n \lambda_l}{k} \dot{C}_l + \lambda_l^2 C_l = f_l.
\end{align} 
Compared to the Stokes drag force model Eq.~\eqref{eq:eom_mode}, the damping coefficient $\eta_s$ is replaced by $\eta_n \lambda_l/k$. 
We can solve this equation by the same method, and we obtain the kinetic energy as 
\begin{align}
\label{eq:ke_normal}
\mathcal{K}_l 
= \frac{B}{2 \eta_n \lambda_l/k} 
= \frac{B}{2 \eta_n} \frac{k}{m \omega_l^2}. 
\end{align}
The law of equipartition of energy is violated, and the low-frequency vibrations are strongly excited in the normal dissipative force model. 
The factor $k/m \omega_l^2$ comes from the effect of weaker damping for softer mode. 

To confirm this result, we also performed MD simulations for the normal dissipative force model. 
The obtained kinetic energy $\mathcal{K}_l$ are shown as the red symbols in Fig.~\ref{fig:energy}. 
Clearly, the simulation results reproduce the theoretical result Eq.~\eqref{eq:ke_normal} well.
\section{\label{sec:sp_structure}Spatial structure of the velocity field}
\begin{figure*}[bt]
  \centering
  \includegraphics[keepaspectratio, width=\linewidth]{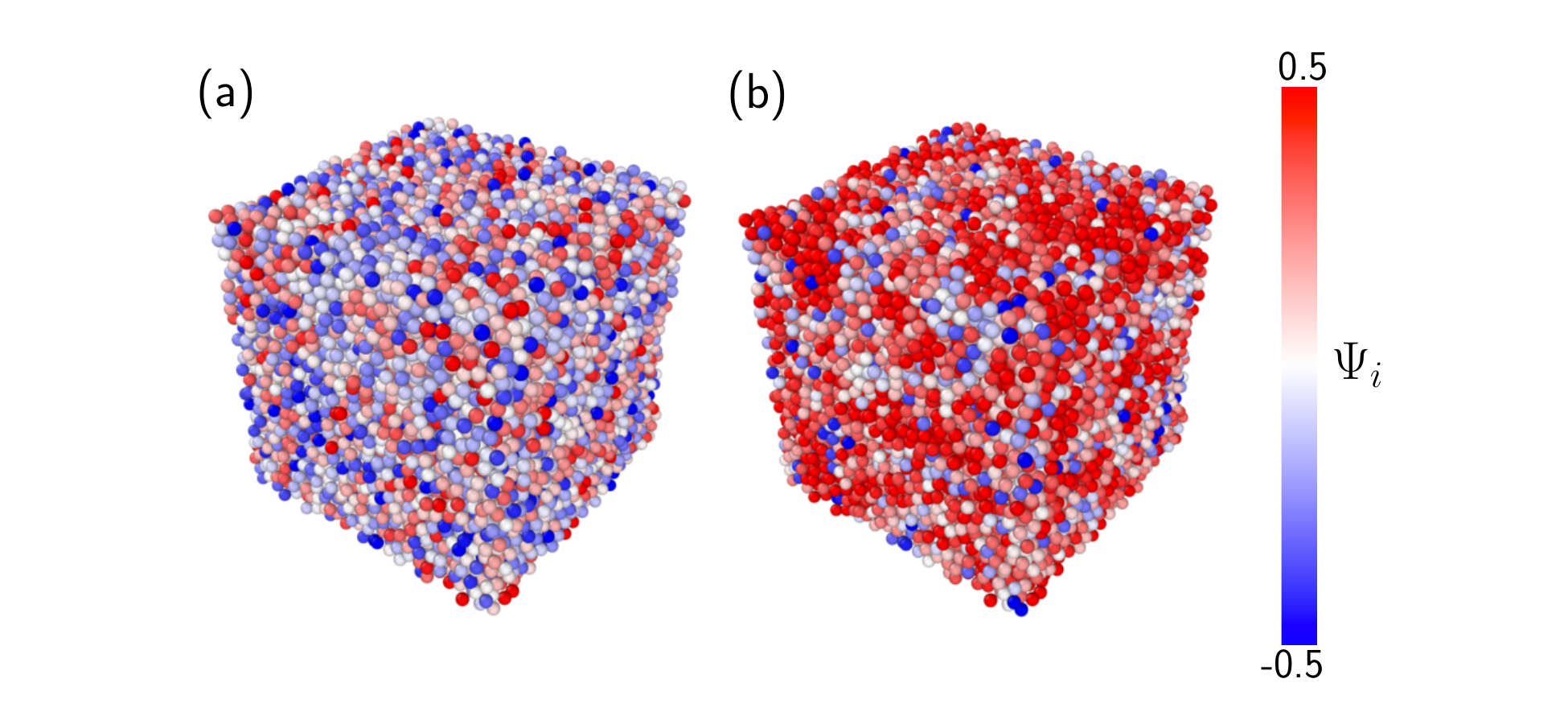}
  \caption{
  Order parameter $\Psi_{i}$ of the granular systems. Visualization of the spatial structure of the vibrations. Results for N=16000 are shown. (a) Stokes drag model, (b) normal dissipative force model. The color of the particles reflects the value of the order parameter, corresponding to the color bar on the right.}\label{fig:op}
\end{figure*}
The previous section showed that the law of equipartition of energy is violated in the normal dissipative force model and the low-frequency vibrations are strongly excited due to the effect of weaker damping for softer modes. 
In this section, we show that this effect causes a collective behavior in the velocity field. 

First, we use MD simulations to calculate $\Psi_{i}$ that quantifies the similarity in velocities of the neighboring particles. 
Figure~\ref{fig:op} shows the snapshots of particles colored by the value of $\Psi_{i}$. 
In the Stokes drag model, $\Psi_{i}$ of each particle is random and there is no noticeable spatial structure. 
On the other hand, in the normal dissipative force model, the particles with larger $\Psi_{i}$ (colored in red) clusters. 
This means that the particle tend to move in the same direction as the neighboring particles and the collective behavior emerges in the velocity field.

\begin{figure}[tb]
  \centering
  \includegraphics[keepaspectratio, width=\linewidth]{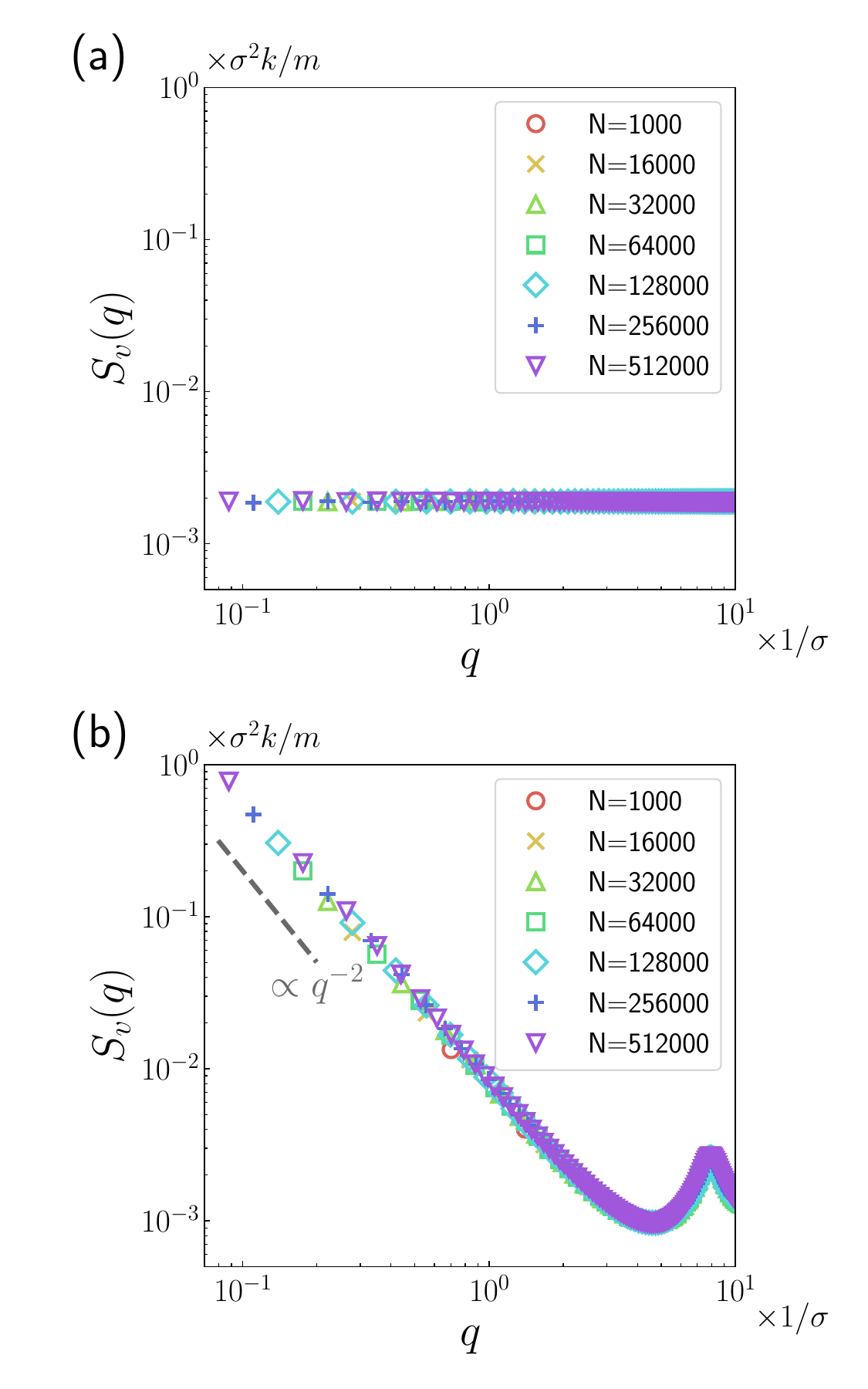}
  \caption{
  Velocity correlation functions obtained from simulations. (a) Stokes drag model (b) Normal dissipative force model.}\label{fig:Sv}
\end{figure}

We quantify this collective behavior by calculating the spatial correlation function of velocities $S_v(q)$. 
Figure~\ref{fig:Sv} shows the results. 
In the Stokes drag model, $S_v(q)$ is constant for all $q$, which suggests a fully random structure of the velocity field. 
This is consistent with the thermal dynamics of this model as shown in the previous section. 
By contrast, in the normal dissipative force model, the correlation function increases and diverges at $q \to 0$. 
In particular, the asymptotic behavior $S_v(q) \propto q^{-2}$ appears in the low-$q$ regime, which suggests the infinitely long-range correlation of the velocities of particles. 

The asymptotic behavior $S_v(q) \propto q^{-2}$ can be understood in terms of the effect of weaker damping for softer mode. 
Here, we focus on the fact that the eigenmodes of the ``unstressed'' hessian matrix $\mathcal{H}$ can be approximated well by the plane waves in the low-frequency regime~\cite{MSI2017}. 
In this approximation, the mode $l$ is characterized by the wave vector $\bm{q}$ and the polarization $\alpha = L, T_1, T_2$, and we write 
\begin{align}
C_l \to C_{\bm{q}, \alpha}, \ \ \ \bm{e}_{l, i} \to \frac{\bm{s}_{\bm{q}, \alpha}}{\sqrt{N}} e^{i \bm{q} \cdot \bm{r}_i}, 
\end{align}
where $\bm{s}_{\bm{k}, \alpha}$ is the normalized polarization vector. 
Then, the Fourier component of the velocity field can be expressed as $\bm{v}_{\bm{q}} \simeq \sum_{\alpha} \dot{C}_{\bm{q}, \alpha} \bm{s}_{\bm{q}, \alpha}$ and the velocity correlation function is obtained as  
\begin{align}
S_v(q) \simeq \sum_{\alpha} \langle \dot{C}_{\bm{q}, \alpha}^2 \rangle. 
\end{align}
This expression should be valid for the low-$q$ regime where the plane wave approximation works. 
In the Stokes drag model, Eq.~\eqref{eq:ke_stokes} leads to $\langle \dot{C}_{\bm{q}, \alpha}^2 \rangle = B/m \eta_s$. 
Therefore, we obtain 
\begin{align}
S_v(q) \simeq \frac{3B}{m \eta_s}. 
\end{align}
This result explains the simulation data for the Stokes drag model very well. 
In the normal dissipative force model, Eq.~\eqref{eq:ke_normal} leads to $\langle \dot{C}_{\bm{q}, \alpha}^2 \rangle = Bk/m^2 \eta_n \omega_{\bm{q}, \alpha}^2$. 
Within the plane wave approximation, the frequency follows the dispersion relation $\omega_{\bm{q}, \alpha} = c_{\alpha} q$, where $c_L$ and $c_T$ is the longitudinal and transverse sound velocity, respectively. 
Then, we obtain 
\begin{align}
S_v(q) \simeq \frac{B}{m \eta_n} \frac{k}{m q^2} \left( \frac{1}{c_L^2} + \frac{2}{c_T^2} \right). 
\end{align}
This result explains the simulation data for the normal dissipative force model where $S_v(q)$ diverges as $q^{-2}$ in the low-$q$ regime. 

This calculation shows the effect of weaker damping for softer mode causes the infinitely long-range velocity correlation in the normal dissipative force model. 
This effect causes the strong excitation for the low-frequency modes as $\langle \dot{C}_l^2 \rangle \propto \omega_l^{-2}$. 
In the low-frequency regime, the eigenmodes are approximated well by the spatially extended plane waves with the dispersion relation $\omega \propto q$, and then these modes are strongly excited as $\langle \dot{C}_{\bm{q}, \alpha}^2 \rangle \propto q^{-2}$. 
This leads to $S_v(q) \propto q^{-2}$. 
\section{\label{sec:conclusion} Conclusion and future works}
In the present work, we examined the vibrational dynamics of random packings of particles agitated by random forces.
We considered two types of models for dissipation: the Stokes drag model, which is appropriate for colloidal particles immersed in a solvent, and the normal dissipative force model, which describes dissipation due to contact between particles and is suitable for granular materials.
By numerically and analytically solving the linearized equation of motion, we demonstrated that the kinetic energy per mode diverges in the low-frequency limit as $\mathcal{K} \propto \omega^{-2}$ in the normal dissipative force model.
This sharply contrasts with the Stokes drag model, where the law of equipartition of energy applies effectively.
This violation of the equipartition law in the normal dissipative force model results from the weaker damping effects for softer modes.
Since the particle displacements parallel to the contacts are minimal in the low-frequency modes, normal dissipation is ineffective in dampening these modes.
Furthermore, we showed that the spatial structure factor of the velocity field exhibits the scaling law $S_v(q) \propto q^{-2}$ in the normal dissipative force model, indicating that the velocity field is infinitely long-ranged.
This also relates to the weaker damping effects for softer modes: low-frequency plane-wave modes are strongly excited, resulting in the emergence of long-range velocity correlations.
These results establish that contact dissipation significantly alters the vibrational dynamics of random packings. 

Our results suggest that the vibrations in granular materials are more collective than those in the thermal glass system.
As mentioned in the introduction section, the vibrational dynamics of granular materials were studied experimentally, and the collective vibrations along with an increase in low-frequency modes were observed~\cite{BDBB2010,HBD2012,OD2013,BWO2024}.
It would be interesting to quantitatively analyze these experimental results based on those obtained in this work.
However, for this purpose, our model remains too simplistic.
The present study considered only the normal dissipative force, while in real experiments, there must be situations where both the normal dissipative force and Stokes drag are involved. Besides, the tangential dissipative force should also be present in granular materials.
To account for this in our model, we need to treat the case where the damping matrix $\Gamma$ and the Hessian matrix $\mathcal{H}$ do not share the same matrix structure.
In this case, the vibrational modes are no longer the eigenvectors of the Hessian matrix $\mathcal{H}$.
Furthermore, the dynamics of real-world granular materials are affected by the friction force arising from the contacts between particles.
Understanding these effects on vibrational dynamics will be the focus of future studies on granular materials.

\section*{Data availability}
The data that support the findings of this study are available from the corresponding author upon reasonable request. Source data are provided with this paper.

\section*{Conflicts of interest}
There are no conflicts of interest to declare.

\begin{acknowledgments}
This work was supported by JSPS KAKENHI Grant Numbers JP20H01868, JP20H00128, JP22K03543, JP23H04495, and JP24H00192.
In this research work, we used the “mdx: a platform for building data-empowered society”.
\end{acknowledgments}

\appendix
\section{Derivation of the kinetic energy}
In both the Stokes drag model and the normal dissipative force model, the linearized equations of motion for each mode have the following form 
\begin{align}
m \ddot{C} + \eta \dot{C} + \lambda C = f, 
\end{align}
where $m, \eta, \lambda$ are constants and $f$ is the random noise satisfying $\langle f^2 \rangle = 2B$. 
This Langevin equation can be solved using the standard method of variation of constants. 

We first transform this equation to the matrix form: 
\begin{align}
\label{eq:langevin_mat}
    \frac{d}{dt}\bm{a}(t) = \Theta \cdot \bm{a}(t) + \bm{F}(t).
\end{align}
where $\bm{a}(t)=(C(t), \dot{C}(t))^\top$, $\bm{F}(t) = (0,f(t)/m)^\top$, and 
\begin{align}
    \Theta=\begin{pmatrix}
            0 & 1 \\
            -\lambda/m & -\eta/m \\
            \end{pmatrix}. 
\end{align}
The covariance matrix of $\bm{F}$ is given by 
\begin{align}
\label{eq:cov_mat_eq}
    \langle \bm{F}(t) \bm{F}^\top(t) \rangle = 2\bm{B}
           = \begin{pmatrix}
             0 & 0 \\
             0 & 2B/m^2 \\
            \end{pmatrix}.
\end{align}
Since this is the first order differential equation, the formal solution can be written as 
\begin{align}
\label{eq:solution_langevin}
\bm{a}(t) = \int_{-\infty}^t ds e^{(t-s)\Theta} \cdot \bm{F}(s).
\end{align}
Because the kinetic energy is given by $\langle \dot{C}^2 \rangle$, we need to evaluate the covariance matrix of $\bm{a}$. 
This can be calculated as 
\begin{align}
\label{eq:cov_mat_eq}
    \langle \bm{a}(t) \bm{a}^\top(t) \rangle 
    &= \int_{- \infty}^t ds e^{(t-s)\Theta} \cdot 2 \bm{B} \cdot e^{(t-s)\Theta^\dagger} \nonumber \\
    &= \int_0^\infty dt e^{t\Theta} \cdot 2 \bm{B} \cdot e^{t\Theta^\dagger},
\end{align}
where $\dagger$ denotes the conjugate transpose. 
By diagonalizing $\Theta$, the matrix $e^{t \Theta}$  and $e^{t \Theta^\dagger}$ can be calculated in a lengthy but straightforward way. 
The obtained result is 
\begin{align}
\label{eq:cov_mat_eq}
 \langle \bm{a}(t) \bm{a}^\top(t) \rangle 
           = \begin{pmatrix}
             \langle C^2 \rangle  & \langle C \dot{C} \rangle \\
             \langle C \dot{C} \rangle & \langle \dot{C}^2 \rangle \\
            \end{pmatrix}
           = \begin{pmatrix}
             B/\eta \lambda  & 0 \\
             0 & B / m \eta \\
            \end{pmatrix}.             
 \end{align}
In the case of the Stokes drag model, the parameter is given by $\eta = \eta_s$ and then 
\begin{align}
\mathcal{K}_l = \frac{1}{2} m \langle \dot{C}_l^2 \rangle = \frac{B}{2 \eta_s}. 
\end{align}
In the case of the normal dissipative force model, the parameter is given by $\eta = \eta_n \lambda_l /k$ and then 
\begin{align}
\mathcal{K}_l = \frac{1}{2} m \langle \dot{C}_l^2 \rangle = \frac{B}{2 \eta_n \lambda_l /k}. 
\end{align}

\bibliography{main_arXiv/main_arXiv}

\begin{thebibliography}{44}%
\makeatletter
\providecommand \@ifxundefined [1]{%
 \@ifx{#1\undefined}
}%
\providecommand \@ifnum [1]{%
 \ifnum #1\expandafter \@firstoftwo
 \else \expandafter \@secondoftwo
 \fi
}%
\providecommand \@ifx [1]{%
 \ifx #1\expandafter \@firstoftwo
 \else \expandafter \@secondoftwo
 \fi
}%
\providecommand \natexlab [1]{#1}%
\providecommand \enquote  [1]{``#1''}%
\providecommand \bibnamefont  [1]{#1}%
\providecommand \bibfnamefont [1]{#1}%
\providecommand \citenamefont [1]{#1}%
\providecommand \href@noop [0]{\@secondoftwo}%
\providecommand \href [0]{\begingroup \@sanitize@url \@href}%
\providecommand \@href[1]{\@@startlink{#1}\@@href}%
\providecommand \@@href[1]{\endgroup#1\@@endlink}%
\providecommand \@sanitize@url [0]{\catcode `\\12\catcode `\$12\catcode `\&12\catcode `\#12\catcode `\^12\catcode `\_12\catcode `\%12\relax}%
\providecommand \@@startlink[1]{}%
\providecommand \@@endlink[0]{}%
\providecommand \url  [0]{\begingroup\@sanitize@url \@url }%
\providecommand \@url [1]{\endgroup\@href {#1}{\urlprefix }}%
\providecommand \urlprefix  [0]{URL }%
\providecommand \Eprint [0]{\href }%
\providecommand \doibase [0]{https://doi.org/}%
\providecommand \selectlanguage [0]{\@gobble}%
\providecommand \bibinfo  [0]{\@secondoftwo}%
\providecommand \bibfield  [0]{\@secondoftwo}%
\providecommand \translation [1]{[#1]}%
\providecommand \BibitemOpen [0]{}%
\providecommand \bibitemStop [0]{}%
\providecommand \bibitemNoStop [0]{.\EOS\space}%
\providecommand \EOS [0]{\spacefactor3000\relax}%
\providecommand \BibitemShut  [1]{\csname bibitem#1\endcsname}%
\let\auto@bib@innerbib\@empty
\bibitem [{\citenamefont {Jaeger}\ and\ \citenamefont {Nagel}(1992)}]{JN1992}%
  \BibitemOpen
  \bibfield  {author} {\bibinfo {author} {\bibfnamefont {H.~M.}\ \bibnamefont {Jaeger}}\ and\ \bibinfo {author} {\bibfnamefont {S.~R.}\ \bibnamefont {Nagel}},\ }\bibfield  {title} {\bibinfo {title} {Physics of the granular state},\ }\href@noop {} {\bibfield  {journal} {\bibinfo  {journal} {Science}\ }\textbf {\bibinfo {volume} {255}},\ \bibinfo {pages} {1523} (\bibinfo {year} {1992})}\BibitemShut {NoStop}%
\bibitem [{\citenamefont {de~Gennes}(1999)}]{deGennes1999}%
  \BibitemOpen
  \bibfield  {author} {\bibinfo {author} {\bibfnamefont {P.~G.}\ \bibnamefont {de~Gennes}},\ }\bibfield  {title} {\bibinfo {title} {Granular matter: a tentative view},\ }\href@noop {} {\bibfield  {journal} {\bibinfo  {journal} {Rev. Mod. Phys.}\ }\textbf {\bibinfo {volume} {71}},\ \bibinfo {pages} {S374} (\bibinfo {year} {1999})}\BibitemShut {NoStop}%
\bibitem [{\citenamefont {Borderies}\ \emph {et~al.}(1985)\citenamefont {Borderies}, \citenamefont {Goldreich},\ and\ \citenamefont {Tremaine}}]{BGT1985}%
  \BibitemOpen
  \bibfield  {author} {\bibinfo {author} {\bibfnamefont {N.}~\bibnamefont {Borderies}}, \bibinfo {author} {\bibfnamefont {P.}~\bibnamefont {Goldreich}},\ and\ \bibinfo {author} {\bibfnamefont {S.}~\bibnamefont {Tremaine}},\ }\bibfield  {title} {\bibinfo {title} {A granular flow model for dense planetary rings},\ }\href@noop {} {\bibfield  {journal} {\bibinfo  {journal} {Icarus}\ }\textbf {\bibinfo {volume} {63}},\ \bibinfo {pages} {406} (\bibinfo {year} {1985})}\BibitemShut {NoStop}%
\bibitem [{\citenamefont {Jaeger}\ \emph {et~al.}(1996)\citenamefont {Jaeger}, \citenamefont {Nagel},\ and\ \citenamefont {Behringer}}]{JNB1996}%
  \BibitemOpen
  \bibfield  {author} {\bibinfo {author} {\bibfnamefont {H.~M.}\ \bibnamefont {Jaeger}}, \bibinfo {author} {\bibfnamefont {S.~R.}\ \bibnamefont {Nagel}},\ and\ \bibinfo {author} {\bibfnamefont {R.~P.}\ \bibnamefont {Behringer}},\ }\bibfield  {title} {\bibinfo {title} {Granular solids, liquids, and gases},\ }\href@noop {} {\bibfield  {journal} {\bibinfo  {journal} {Rev. Mod. Phys.}\ }\textbf {\bibinfo {volume} {68}},\ \bibinfo {pages} {1259} (\bibinfo {year} {1996})}\BibitemShut {NoStop}%
\bibitem [{\citenamefont {Daerr}\ and\ \citenamefont {Douady}(1999)}]{DD1999}%
  \BibitemOpen
  \bibfield  {author} {\bibinfo {author} {\bibfnamefont {A.}~\bibnamefont {Daerr}}\ and\ \bibinfo {author} {\bibfnamefont {S.}~\bibnamefont {Douady}},\ }\bibfield  {title} {\bibinfo {title} {Two types of avalanche behaviour in granular media},\ }\href@noop {} {\bibfield  {journal} {\bibinfo  {journal} {Nature}\ }\textbf {\bibinfo {volume} {399}},\ \bibinfo {pages} {241} (\bibinfo {year} {1999})}\BibitemShut {NoStop}%
\bibitem [{\citenamefont {Aranson}\ and\ \citenamefont {Tsimring}(2006)}]{AT2006}%
  \BibitemOpen
  \bibfield  {author} {\bibinfo {author} {\bibfnamefont {I.~S.}\ \bibnamefont {Aranson}}\ and\ \bibinfo {author} {\bibfnamefont {L.~S.}\ \bibnamefont {Tsimring}},\ }\bibfield  {title} {\bibinfo {title} {Patterns and collective behavior in granular media: Theoretical concepts},\ }\href@noop {} {\bibfield  {journal} {\bibinfo  {journal} {Rev. Mod. Phys.}\ }\textbf {\bibinfo {volume} {78}},\ \bibinfo {pages} {641} (\bibinfo {year} {2006})}\BibitemShut {NoStop}%
\bibitem [{\citenamefont {Goldhirsch}\ and\ \citenamefont {Zanetti}(1993)}]{GZ1993}%
  \BibitemOpen
  \bibfield  {author} {\bibinfo {author} {\bibfnamefont {I.}~\bibnamefont {Goldhirsch}}\ and\ \bibinfo {author} {\bibfnamefont {G.}~\bibnamefont {Zanetti}},\ }\bibfield  {title} {\bibinfo {title} {Clustering instability in dissipative gases},\ }\href {https://doi.org/10.1103/PhysRevLett.70.1619} {\bibfield  {journal} {\bibinfo  {journal} {Phys. Rev. Lett.}\ }\textbf {\bibinfo {volume} {70}},\ \bibinfo {pages} {1619} (\bibinfo {year} {1993})}\BibitemShut {NoStop}%
\bibitem [{\citenamefont {Pathak}\ \emph {et~al.}(2014)\citenamefont {Pathak}, \citenamefont {Jabeen}, \citenamefont {Das},\ and\ \citenamefont {Rajesh}}]{PJD2014}%
  \BibitemOpen
  \bibfield  {author} {\bibinfo {author} {\bibfnamefont {S.~N.}\ \bibnamefont {Pathak}}, \bibinfo {author} {\bibfnamefont {Z.}~\bibnamefont {Jabeen}}, \bibinfo {author} {\bibfnamefont {D.}~\bibnamefont {Das}},\ and\ \bibinfo {author} {\bibfnamefont {R.}~\bibnamefont {Rajesh}},\ }\bibfield  {title} {\bibinfo {title} {Energy decay in three-dimensional freely cooling granular gas},\ }\href {https://doi.org/10.1103/PhysRevLett.112.038001} {\bibfield  {journal} {\bibinfo  {journal} {Phys. Rev. Lett.}\ }\textbf {\bibinfo {volume} {112}},\ \bibinfo {pages} {038001} (\bibinfo {year} {2014})}\BibitemShut {NoStop}%
\bibitem [{\citenamefont {Gallas}\ \emph {et~al.}(1992)\citenamefont {Gallas}, \citenamefont {Herrmann},\ and\ \citenamefont {Soko{\l}owski}}]{GHS1992}%
  \BibitemOpen
  \bibfield  {author} {\bibinfo {author} {\bibfnamefont {J.~A.~C.}\ \bibnamefont {Gallas}}, \bibinfo {author} {\bibfnamefont {H.~J.}\ \bibnamefont {Herrmann}},\ and\ \bibinfo {author} {\bibfnamefont {S.}~\bibnamefont {Soko{\l}owski}},\ }\bibfield  {title} {\bibinfo {title} {Convection cells in vibrating granular media},\ }\href@noop {} {\bibfield  {journal} {\bibinfo  {journal} {Phys. Rev. Lett.}\ }\textbf {\bibinfo {volume} {69}},\ \bibinfo {pages} {1371} (\bibinfo {year} {1992})}\BibitemShut {NoStop}%
\bibitem [{\citenamefont {Conway}\ \emph {et~al.}(2004)\citenamefont {Conway}, \citenamefont {Shinbrot},\ and\ \citenamefont {Glasser}}]{CSG2004}%
  \BibitemOpen
  \bibfield  {author} {\bibinfo {author} {\bibfnamefont {S.~L.}\ \bibnamefont {Conway}}, \bibinfo {author} {\bibfnamefont {T.}~\bibnamefont {Shinbrot}},\ and\ \bibinfo {author} {\bibfnamefont {B.~J.}\ \bibnamefont {Glasser}},\ }\bibfield  {title} {\bibinfo {title} {A taylor vortex analogy in granular flows},\ }\href@noop {} {\bibfield  {journal} {\bibinfo  {journal} {Nature}\ }\textbf {\bibinfo {volume} {431}},\ \bibinfo {pages} {433} (\bibinfo {year} {2004})}\BibitemShut {NoStop}%
\bibitem [{\citenamefont {Jop}\ \emph {et~al.}(2006)\citenamefont {Jop}, \citenamefont {Forterre},\ and\ \citenamefont {Pouliquen}}]{JFP2006}%
  \BibitemOpen
  \bibfield  {author} {\bibinfo {author} {\bibfnamefont {P.}~\bibnamefont {Jop}}, \bibinfo {author} {\bibfnamefont {Y.}~\bibnamefont {Forterre}},\ and\ \bibinfo {author} {\bibfnamefont {O.}~\bibnamefont {Pouliquen}},\ }\bibfield  {title} {\bibinfo {title} {A constitutive law for dense granular flows},\ }\href@noop {} {\bibfield  {journal} {\bibinfo  {journal} {Nature}\ }\textbf {\bibinfo {volume} {441}},\ \bibinfo {pages} {727} (\bibinfo {year} {2006})}\BibitemShut {NoStop}%
\bibitem [{\citenamefont {Abedi}\ \emph {et~al.}(2012)\citenamefont {Abedi}, \citenamefont {Rechenmacher},\ and\ \citenamefont {Orlando}}]{ARO2012}%
  \BibitemOpen
  \bibfield  {author} {\bibinfo {author} {\bibfnamefont {S.}~\bibnamefont {Abedi}}, \bibinfo {author} {\bibfnamefont {A.~L.}\ \bibnamefont {Rechenmacher}},\ and\ \bibinfo {author} {\bibfnamefont {A.~D.}\ \bibnamefont {Orlando}},\ }\bibfield  {title} {\bibinfo {title} {Vortex formation and dissolution in sheared sands},\ }\href@noop {} {\bibfield  {journal} {\bibinfo  {journal} {Granul. Matter}\ }\textbf {\bibinfo {volume} {14}},\ \bibinfo {pages} {695} (\bibinfo {year} {2012})}\BibitemShut {NoStop}%
\bibitem [{\citenamefont {Fall}\ \emph {et~al.}(2015)\citenamefont {Fall}, \citenamefont {Ovarlez}, \citenamefont {Hautemayou}, \citenamefont {M{\'e}zi{\`e}re}, \citenamefont {Roux},\ and\ \citenamefont {Chevoir}}]{FOH+2015}%
  \BibitemOpen
  \bibfield  {author} {\bibinfo {author} {\bibfnamefont {A.}~\bibnamefont {Fall}}, \bibinfo {author} {\bibfnamefont {G.}~\bibnamefont {Ovarlez}}, \bibinfo {author} {\bibfnamefont {D.}~\bibnamefont {Hautemayou}}, \bibinfo {author} {\bibfnamefont {C.}~\bibnamefont {M{\'e}zi{\`e}re}}, \bibinfo {author} {\bibfnamefont {J.-N.}\ \bibnamefont {Roux}},\ and\ \bibinfo {author} {\bibfnamefont {F.}~\bibnamefont {Chevoir}},\ }\bibfield  {title} {\bibinfo {title} {Dry granular flows: Rheological measurements of the $\mu$ (i)-rheology},\ }\href@noop {} {\bibfield  {journal} {\bibinfo  {journal} {J. Rheol.}\ }\textbf {\bibinfo {volume} {59}},\ \bibinfo {pages} {1065} (\bibinfo {year} {2015})}\BibitemShut {NoStop}%
\bibitem [{\citenamefont {Krishnaraj}\ and\ \citenamefont {Nott}(2016)}]{KN2016}%
  \BibitemOpen
  \bibfield  {author} {\bibinfo {author} {\bibfnamefont {K.~P.}\ \bibnamefont {Krishnaraj}}\ and\ \bibinfo {author} {\bibfnamefont {P.~R.}\ \bibnamefont {Nott}},\ }\bibfield  {title} {\bibinfo {title} {A dilation-driven vortex flow in sheared granular materials explains a rheometric anomaly},\ }\href@noop {} {\bibfield  {journal} {\bibinfo  {journal} {Nat. Commun.}\ }\textbf {\bibinfo {volume} {7}},\ \bibinfo {pages} {10630} (\bibinfo {year} {2016})}\BibitemShut {NoStop}%
\bibitem [{\citenamefont {Windows-Yule}(2016)}]{Windows2016}%
  \BibitemOpen
  \bibfield  {author} {\bibinfo {author} {\bibfnamefont {C.~R.~K.}\ \bibnamefont {Windows-Yule}},\ }\bibfield  {title} {\bibinfo {title} {Convection and segregation in fluidised granular systems exposed to two-dimensional vibration},\ }\href@noop {} {\bibfield  {journal} {\bibinfo  {journal} {New J. Phys.}\ }\textbf {\bibinfo {volume} {18}},\ \bibinfo {pages} {033005} (\bibinfo {year} {2016})}\BibitemShut {NoStop}%
\bibitem [{\citenamefont {Kharel}\ and\ \citenamefont {Rognon}(2017)}]{KR2017}%
  \BibitemOpen
  \bibfield  {author} {\bibinfo {author} {\bibfnamefont {P.}~\bibnamefont {Kharel}}\ and\ \bibinfo {author} {\bibfnamefont {P.}~\bibnamefont {Rognon}},\ }\bibfield  {title} {\bibinfo {title} {Vortices enhance diffusion in dense granular flows},\ }\href@noop {} {\bibfield  {journal} {\bibinfo  {journal} {Phys. Rev. Lett.}\ }\textbf {\bibinfo {volume} {119}},\ \bibinfo {pages} {178001} (\bibinfo {year} {2017})}\BibitemShut {NoStop}%
\bibitem [{\citenamefont {Kozicki}\ and\ \citenamefont {Tejchman}(2018)}]{KT2018}%
  \BibitemOpen
  \bibfield  {author} {\bibinfo {author} {\bibfnamefont {J.}~\bibnamefont {Kozicki}}\ and\ \bibinfo {author} {\bibfnamefont {J.}~\bibnamefont {Tejchman}},\ }\bibfield  {title} {\bibinfo {title} {Relationship between vortex structures and shear localization in 3d granular specimens based on combined dem and helmholtz--hodge decomposition},\ }\href@noop {} {\bibfield  {journal} {\bibinfo  {journal} {Granul. Matter}\ }\textbf {\bibinfo {volume} {20}},\ \bibinfo {pages} {48} (\bibinfo {year} {2018})}\BibitemShut {NoStop}%
\bibitem [{\citenamefont {Ahmad}\ and\ \citenamefont {Smalley}(1973)}]{AS1973}%
  \BibitemOpen
  \bibfield  {author} {\bibinfo {author} {\bibfnamefont {K.}~\bibnamefont {Ahmad}}\ and\ \bibinfo {author} {\bibfnamefont {I.~J.}\ \bibnamefont {Smalley}},\ }\bibfield  {title} {\bibinfo {title} {Observation of particle segregation in vibrated granular systems},\ }\href@noop {} {\bibfield  {journal} {\bibinfo  {journal} {Powder Technol.}\ }\textbf {\bibinfo {volume} {8}},\ \bibinfo {pages} {69} (\bibinfo {year} {1973})}\BibitemShut {NoStop}%
\bibitem [{\citenamefont {O'Hern}\ \emph {et~al.}(2003)\citenamefont {O'Hern}, \citenamefont {Silbert}, \citenamefont {Liu},\ and\ \citenamefont {Nagel}}]{OSL+2003}%
  \BibitemOpen
  \bibfield  {author} {\bibinfo {author} {\bibfnamefont {C.~S.}\ \bibnamefont {O'Hern}}, \bibinfo {author} {\bibfnamefont {L.~E.}\ \bibnamefont {Silbert}}, \bibinfo {author} {\bibfnamefont {A.~J.}\ \bibnamefont {Liu}},\ and\ \bibinfo {author} {\bibfnamefont {S.~R.}\ \bibnamefont {Nagel}},\ }\bibfield  {title} {\bibinfo {title} {Jamming at zero temperature and zero applied stress: The epitome of disorder},\ }\href@noop {} {\bibfield  {journal} {\bibinfo  {journal} {Phys. Rev. E}\ }\textbf {\bibinfo {volume} {68}},\ \bibinfo {pages} {011306} (\bibinfo {year} {2003})}\BibitemShut {NoStop}%
\bibitem [{\citenamefont {Princen}\ and\ \citenamefont {Kiss}(1986)}]{PK1986}%
  \BibitemOpen
  \bibfield  {author} {\bibinfo {author} {\bibfnamefont {H.~M.}\ \bibnamefont {Princen}}\ and\ \bibinfo {author} {\bibfnamefont {A.~D.}\ \bibnamefont {Kiss}},\ }\bibfield  {title} {\bibinfo {title} {Rheology of foams and highly concentrated emulsions: Iii. static shear modulus},\ }\href@noop {} {\bibfield  {journal} {\bibinfo  {journal} {J. Colloid Interface Sci.}\ }\textbf {\bibinfo {volume} {112}},\ \bibinfo {pages} {427} (\bibinfo {year} {1986})}\BibitemShut {NoStop}%
\bibitem [{\citenamefont {Bolton}\ and\ \citenamefont {Weaire}(1990)}]{BW1990}%
  \BibitemOpen
  \bibfield  {author} {\bibinfo {author} {\bibfnamefont {F.}~\bibnamefont {Bolton}}\ and\ \bibinfo {author} {\bibfnamefont {D.}~\bibnamefont {Weaire}},\ }\bibfield  {title} {\bibinfo {title} {Rigidity loss transition in a disordered 2d froth},\ }\href@noop {} {\bibfield  {journal} {\bibinfo  {journal} {Phys. Rev. Lett.}\ }\textbf {\bibinfo {volume} {65}},\ \bibinfo {pages} {3449} (\bibinfo {year} {1990})}\BibitemShut {NoStop}%
\bibitem [{\citenamefont {Majmudar}\ \emph {et~al.}(2007)\citenamefont {Majmudar}, \citenamefont {Sperl}, \citenamefont {Luding},\ and\ \citenamefont {Behringer}}]{MSL+2007}%
  \BibitemOpen
  \bibfield  {author} {\bibinfo {author} {\bibfnamefont {T.~S.}\ \bibnamefont {Majmudar}}, \bibinfo {author} {\bibfnamefont {M.}~\bibnamefont {Sperl}}, \bibinfo {author} {\bibfnamefont {S.}~\bibnamefont {Luding}},\ and\ \bibinfo {author} {\bibfnamefont {R.~P.}\ \bibnamefont {Behringer}},\ }\bibfield  {title} {\bibinfo {title} {Jamming transition in granular systems},\ }\href@noop {} {\bibfield  {journal} {\bibinfo  {journal} {Phys. Rev. Lett.}\ }\textbf {\bibinfo {volume} {98}},\ \bibinfo {pages} {058001} (\bibinfo {year} {2007})}\BibitemShut {NoStop}%
\bibitem [{\citenamefont {van Hecke}(2009)}]{vanHecke2009}%
  \BibitemOpen
  \bibfield  {author} {\bibinfo {author} {\bibfnamefont {M.}~\bibnamefont {van Hecke}},\ }\bibfield  {title} {\bibinfo {title} {Jamming of soft particles: geometry, mechanics, scaling and isostaticity},\ }\href@noop {} {\bibfield  {journal} {\bibinfo  {journal} {J. Phys. Condens. Matter}\ }\textbf {\bibinfo {volume} {22}},\ \bibinfo {pages} {033101} (\bibinfo {year} {2009})}\BibitemShut {NoStop}%
\bibitem [{\citenamefont {Clusel}\ \emph {et~al.}(2009)\citenamefont {Clusel}, \citenamefont {Corwin}, \citenamefont {Siemens},\ and\ \citenamefont {Bruji{\'c}}}]{CCS+2009}%
  \BibitemOpen
  \bibfield  {author} {\bibinfo {author} {\bibfnamefont {M.}~\bibnamefont {Clusel}}, \bibinfo {author} {\bibfnamefont {E.~I.}\ \bibnamefont {Corwin}}, \bibinfo {author} {\bibfnamefont {A.~O.~N.}\ \bibnamefont {Siemens}},\ and\ \bibinfo {author} {\bibfnamefont {J.}~\bibnamefont {Bruji{\'c}}},\ }\bibfield  {title} {\bibinfo {title} {A 'granocentric' model for random packing of jammed emulsions},\ }\href@noop {} {\bibfield  {journal} {\bibinfo  {journal} {Nature}\ }\textbf {\bibinfo {volume} {460}},\ \bibinfo {pages} {611} (\bibinfo {year} {2009})}\BibitemShut {NoStop}%
\bibitem [{\citenamefont {Brown}\ \emph {et~al.}(2010)\citenamefont {Brown}, \citenamefont {Rodenberg}, \citenamefont {Amend}, \citenamefont {Mozeika}, \citenamefont {Steltz}, \citenamefont {Zakin}, \citenamefont {Lipson},\ and\ \citenamefont {Jaeger}}]{BRA+2010}%
  \BibitemOpen
  \bibfield  {author} {\bibinfo {author} {\bibfnamefont {E.}~\bibnamefont {Brown}}, \bibinfo {author} {\bibfnamefont {N.}~\bibnamefont {Rodenberg}}, \bibinfo {author} {\bibfnamefont {J.}~\bibnamefont {Amend}}, \bibinfo {author} {\bibfnamefont {A.}~\bibnamefont {Mozeika}}, \bibinfo {author} {\bibfnamefont {E.}~\bibnamefont {Steltz}}, \bibinfo {author} {\bibfnamefont {M.~R.}\ \bibnamefont {Zakin}}, \bibinfo {author} {\bibfnamefont {H.}~\bibnamefont {Lipson}},\ and\ \bibinfo {author} {\bibfnamefont {H.~M.}\ \bibnamefont {Jaeger}},\ }\bibfield  {title} {\bibinfo {title} {Universal robotic gripper based on the jamming of granular material},\ }\href@noop {} {\bibfield  {journal} {\bibinfo  {journal} {Proc. Natl. Acad. Sci. USA}\ }\textbf {\bibinfo {volume} {107}},\ \bibinfo {pages} {18809} (\bibinfo {year} {2010})}\BibitemShut {NoStop}%
\bibitem [{\citenamefont {Behringer}\ and\ \citenamefont {Chakraborty}(2018)}]{BC2018}%
  \BibitemOpen
  \bibfield  {author} {\bibinfo {author} {\bibfnamefont {R.~P.}\ \bibnamefont {Behringer}}\ and\ \bibinfo {author} {\bibfnamefont {B.}~\bibnamefont {Chakraborty}},\ }\bibfield  {title} {\bibinfo {title} {The physics of jamming for granular materials: a review},\ }\href@noop {} {\bibfield  {journal} {\bibinfo  {journal} {Rep. Prog. Phys.}\ }\textbf {\bibinfo {volume} {82}},\ \bibinfo {pages} {012601} (\bibinfo {year} {2018})}\BibitemShut {NoStop}%
\bibitem [{\citenamefont {O'Hern}\ \emph {et~al.}(2002)\citenamefont {O'Hern}, \citenamefont {Langer}, \citenamefont {Liu},\ and\ \citenamefont {Nagel}}]{OLL+2002}%
  \BibitemOpen
  \bibfield  {author} {\bibinfo {author} {\bibfnamefont {C.~S.}\ \bibnamefont {O'Hern}}, \bibinfo {author} {\bibfnamefont {S.~A.}\ \bibnamefont {Langer}}, \bibinfo {author} {\bibfnamefont {A.~J.}\ \bibnamefont {Liu}},\ and\ \bibinfo {author} {\bibfnamefont {S.~R.}\ \bibnamefont {Nagel}},\ }\bibfield  {title} {\bibinfo {title} {Random packings of frictionless particles},\ }\href@noop {} {\bibfield  {journal} {\bibinfo  {journal} {Phys. Rev. Lett.}\ }\textbf {\bibinfo {volume} {88}},\ \bibinfo {pages} {075507} (\bibinfo {year} {2002})}\BibitemShut {NoStop}%
\bibitem [{\citenamefont {Mizuno}\ \emph {et~al.}(2017)\citenamefont {Mizuno}, \citenamefont {Shiba},\ and\ \citenamefont {Ikeda}}]{MSI2017}%
  \BibitemOpen
  \bibfield  {author} {\bibinfo {author} {\bibfnamefont {H.}~\bibnamefont {Mizuno}}, \bibinfo {author} {\bibfnamefont {H.}~\bibnamefont {Shiba}},\ and\ \bibinfo {author} {\bibfnamefont {A.}~\bibnamefont {Ikeda}},\ }\bibfield  {title} {\bibinfo {title} {Continuum limit of the vibrational properties of amorphous solids},\ }\href {https://doi.org/10.1073/pnas.1709015114} {\bibfield  {journal} {\bibinfo  {journal} {Proc. Natl. Acad. Sci. USA}\ }\textbf {\bibinfo {volume} {114}},\ \bibinfo {pages} {E9767} (\bibinfo {year} {2017})}\BibitemShut {NoStop}%
\bibitem [{\citenamefont {Kapteijns}\ \emph {et~al.}(2018)\citenamefont {Kapteijns}, \citenamefont {Bouchbinder},\ and\ \citenamefont {Lerner}}]{KBL2018}%
  \BibitemOpen
  \bibfield  {author} {\bibinfo {author} {\bibfnamefont {G.}~\bibnamefont {Kapteijns}}, \bibinfo {author} {\bibfnamefont {E.}~\bibnamefont {Bouchbinder}},\ and\ \bibinfo {author} {\bibfnamefont {E.}~\bibnamefont {Lerner}},\ }\bibfield  {title} {\bibinfo {title} {Universal nonphononic density of states in 2d, 3d, and 4d glasses},\ }\href@noop {} {\bibfield  {journal} {\bibinfo  {journal} {Phys. Rev. Lett.}\ }\textbf {\bibinfo {volume} {121}},\ \bibinfo {pages} {055501} (\bibinfo {year} {2018})}\BibitemShut {NoStop}%
\bibitem [{\citenamefont {Shimada}\ \emph {et~al.}(2018)\citenamefont {Shimada}, \citenamefont {Mizuno},\ and\ \citenamefont {Ikeda}}]{SMI2018}%
  \BibitemOpen
  \bibfield  {author} {\bibinfo {author} {\bibfnamefont {M.}~\bibnamefont {Shimada}}, \bibinfo {author} {\bibfnamefont {H.}~\bibnamefont {Mizuno}},\ and\ \bibinfo {author} {\bibfnamefont {A.}~\bibnamefont {Ikeda}},\ }\bibfield  {title} {\bibinfo {title} {Anomalous vibrational properties in the continuum limit of glasses},\ }\href@noop {} {\bibfield  {journal} {\bibinfo  {journal} {Phys. Rev. E}\ }\textbf {\bibinfo {volume} {97}},\ \bibinfo {pages} {022609} (\bibinfo {year} {2018})}\BibitemShut {NoStop}%
\bibitem [{\citenamefont {Lema{\^\i}tre}\ and\ \citenamefont {Maloney}(2006)}]{LM2006}%
  \BibitemOpen
  \bibfield  {author} {\bibinfo {author} {\bibfnamefont {A.}~\bibnamefont {Lema{\^\i}tre}}\ and\ \bibinfo {author} {\bibfnamefont {C.}~\bibnamefont {Maloney}},\ }\bibfield  {title} {\bibinfo {title} {Sum rules for the quasi-static and visco-elastic response of disordered solids at zero temperature},\ }\href@noop {} {\bibfield  {journal} {\bibinfo  {journal} {J. Stat. Phys.}\ }\textbf {\bibinfo {volume} {123}},\ \bibinfo {pages} {415} (\bibinfo {year} {2006})}\BibitemShut {NoStop}%
\bibitem [{\citenamefont {Brito}\ \emph {et~al.}(2010)\citenamefont {Brito}, \citenamefont {Dauchot}, \citenamefont {Biroli},\ and\ \citenamefont {Bouchaud}}]{BDBB2010}%
  \BibitemOpen
  \bibfield  {author} {\bibinfo {author} {\bibfnamefont {C.}~\bibnamefont {Brito}}, \bibinfo {author} {\bibfnamefont {O.}~\bibnamefont {Dauchot}}, \bibinfo {author} {\bibfnamefont {G.}~\bibnamefont {Biroli}},\ and\ \bibinfo {author} {\bibfnamefont {J.-P.}\ \bibnamefont {Bouchaud}},\ }\bibfield  {title} {\bibinfo {title} {Elementary excitation modes in a granular glass above jamming},\ }\href {https://doi.org/10.1039/C001360A} {\bibfield  {journal} {\bibinfo  {journal} {Soft Matter}\ }\textbf {\bibinfo {volume} {6}},\ \bibinfo {pages} {3013} (\bibinfo {year} {2010})}\BibitemShut {NoStop}%
\bibitem [{\citenamefont {Henkes}\ \emph {et~al.}(2012)\citenamefont {Henkes}, \citenamefont {Brito},\ and\ \citenamefont {Dauchot}}]{HBD2012}%
  \BibitemOpen
  \bibfield  {author} {\bibinfo {author} {\bibfnamefont {S.}~\bibnamefont {Henkes}}, \bibinfo {author} {\bibfnamefont {C.}~\bibnamefont {Brito}},\ and\ \bibinfo {author} {\bibfnamefont {O.}~\bibnamefont {Dauchot}},\ }\bibfield  {title} {\bibinfo {title} {Extracting vibrational modes from fluctuations: a pedagogical discussion},\ }\href@noop {} {\bibfield  {journal} {\bibinfo  {journal} {Soft Matter}\ }\textbf {\bibinfo {volume} {8}},\ \bibinfo {pages} {6092} (\bibinfo {year} {2012})}\BibitemShut {NoStop}%
\bibitem [{\citenamefont {Owens}\ and\ \citenamefont {Daniels}(2013)}]{OD2013}%
  \BibitemOpen
  \bibfield  {author} {\bibinfo {author} {\bibfnamefont {E.~T.}\ \bibnamefont {Owens}}\ and\ \bibinfo {author} {\bibfnamefont {K.~E.}\ \bibnamefont {Daniels}},\ }\bibfield  {title} {\bibinfo {title} {Acoustic measurement of a granular density of modes},\ }\href {https://doi.org/10.1039/C2SM27122B} {\bibfield  {journal} {\bibinfo  {journal} {Soft Matter}\ }\textbf {\bibinfo {volume} {9}},\ \bibinfo {pages} {1214} (\bibinfo {year} {2013})}\BibitemShut {NoStop}%
\bibitem [{\citenamefont {Blue}\ \emph {et~al.}(2024)\citenamefont {Blue}, \citenamefont {Wright},\ and\ \citenamefont {Owens}}]{BWO2024}%
  \BibitemOpen
  \bibfield  {author} {\bibinfo {author} {\bibfnamefont {S.~A.}\ \bibnamefont {Blue}}, \bibinfo {author} {\bibfnamefont {S.~C.}\ \bibnamefont {Wright}},\ and\ \bibinfo {author} {\bibfnamefont {E.~T.}\ \bibnamefont {Owens}},\ }\bibfield  {title} {\bibinfo {title} {Experimental measurements of the granular density of modes via impact},\ }\href {https://doi.org/10.1103/PhysRevE.110.014902} {\bibfield  {journal} {\bibinfo  {journal} {Phys. Rev. E}\ }\textbf {\bibinfo {volume} {110}},\ \bibinfo {pages} {014902} (\bibinfo {year} {2024})}\BibitemShut {NoStop}%
\bibitem [{\citenamefont {Somfai}\ \emph {et~al.}(2007)\citenamefont {Somfai}, \citenamefont {van Hecke}, \citenamefont {Ellenbroek}, \citenamefont {Shundyak},\ and\ \citenamefont {van Saarloos}}]{SvHE+2007}%
  \BibitemOpen
  \bibfield  {author} {\bibinfo {author} {\bibfnamefont {E.}~\bibnamefont {Somfai}}, \bibinfo {author} {\bibfnamefont {M.}~\bibnamefont {van Hecke}}, \bibinfo {author} {\bibfnamefont {W.~G.}\ \bibnamefont {Ellenbroek}}, \bibinfo {author} {\bibfnamefont {K.}~\bibnamefont {Shundyak}},\ and\ \bibinfo {author} {\bibfnamefont {W.}~\bibnamefont {van Saarloos}},\ }\bibfield  {title} {\bibinfo {title} {Critical and noncritical jamming of frictional grains},\ }\href@noop {} {\bibfield  {journal} {\bibinfo  {journal} {Phys. Rev. E: Stat., Nonlinear, Soft Matter Phys.}\ }\textbf {\bibinfo {volume} {75}},\ \bibinfo {pages} {020301} (\bibinfo {year} {2007})}\BibitemShut {NoStop}%
\bibitem [{\citenamefont {Saitoh}\ \emph {et~al.}(2019)\citenamefont {Saitoh}, \citenamefont {Shrivastava},\ and\ \citenamefont {Luding}}]{SSL2019}%
  \BibitemOpen
  \bibfield  {author} {\bibinfo {author} {\bibfnamefont {K.}~\bibnamefont {Saitoh}}, \bibinfo {author} {\bibfnamefont {R.~K.}\ \bibnamefont {Shrivastava}},\ and\ \bibinfo {author} {\bibfnamefont {S.}~\bibnamefont {Luding}},\ }\bibfield  {title} {\bibinfo {title} {Rotational sound in disordered granular materials},\ }\href@noop {} {\bibfield  {journal} {\bibinfo  {journal} {Phys. Rev. E}\ }\textbf {\bibinfo {volume} {99}},\ \bibinfo {pages} {012906} (\bibinfo {year} {2019})}\BibitemShut {NoStop}%
\bibitem [{\citenamefont {Chattoraj}\ \emph {et~al.}(2019)\citenamefont {Chattoraj}, \citenamefont {Gendelman}, \citenamefont {Pica~Ciamarra},\ and\ \citenamefont {Procaccia}}]{CGP+2019}%
  \BibitemOpen
  \bibfield  {author} {\bibinfo {author} {\bibfnamefont {J.}~\bibnamefont {Chattoraj}}, \bibinfo {author} {\bibfnamefont {O.}~\bibnamefont {Gendelman}}, \bibinfo {author} {\bibfnamefont {M.}~\bibnamefont {Pica~Ciamarra}},\ and\ \bibinfo {author} {\bibfnamefont {I.}~\bibnamefont {Procaccia}},\ }\bibfield  {title} {\bibinfo {title} {Oscillatory instabilities in frictional granular matter},\ }\href@noop {} {\bibfield  {journal} {\bibinfo  {journal} {Phys. Rev. Lett.}\ }\textbf {\bibinfo {volume} {123}},\ \bibinfo {pages} {098003} (\bibinfo {year} {2019})}\BibitemShut {NoStop}%
\bibitem [{\citenamefont {Ishima}\ \emph {et~al.}(2023)\citenamefont {Ishima}, \citenamefont {Saitoh}, \citenamefont {Otsuki},\ and\ \citenamefont {Hayakawa}}]{ISO+2023}%
  \BibitemOpen
  \bibfield  {author} {\bibinfo {author} {\bibfnamefont {D.}~\bibnamefont {Ishima}}, \bibinfo {author} {\bibfnamefont {K.}~\bibnamefont {Saitoh}}, \bibinfo {author} {\bibfnamefont {M.}~\bibnamefont {Otsuki}},\ and\ \bibinfo {author} {\bibfnamefont {H.}~\bibnamefont {Hayakawa}},\ }\bibfield  {title} {\bibinfo {title} {Eigenvalue analysis of stress-strain curve of two-dimensional amorphous solids of dispersed frictional grains with finite shear strain},\ }\href@noop {} {\bibfield  {journal} {\bibinfo  {journal} {Phys. Rev. E}\ }\textbf {\bibinfo {volume} {107}},\ \bibinfo {pages} {034904} (\bibinfo {year} {2023})}\BibitemShut {NoStop}%
\bibitem [{\citenamefont {Durian}(1995)}]{Durian1995}%
  \BibitemOpen
  \bibfield  {author} {\bibinfo {author} {\bibfnamefont {D.~J.}\ \bibnamefont {Durian}},\ }\bibfield  {title} {\bibinfo {title} {Foam mechanics at the bubble scale},\ }\href@noop {} {\bibfield  {journal} {\bibinfo  {journal} {Phys. Rev. Lett.}\ }\textbf {\bibinfo {volume} {75}},\ \bibinfo {pages} {4780} (\bibinfo {year} {1995})}\BibitemShut {NoStop}%
\bibitem [{\citenamefont {Henkes}\ \emph {et~al.}(2020)\citenamefont {Henkes}, \citenamefont {Kostanjevec}, \citenamefont {Collinson}, \citenamefont {Sknepnek},\ and\ \citenamefont {Bertin}}]{HKC+2020}%
  \BibitemOpen
  \bibfield  {author} {\bibinfo {author} {\bibfnamefont {S.}~\bibnamefont {Henkes}}, \bibinfo {author} {\bibfnamefont {K.}~\bibnamefont {Kostanjevec}}, \bibinfo {author} {\bibfnamefont {J.~M.}\ \bibnamefont {Collinson}}, \bibinfo {author} {\bibfnamefont {R.}~\bibnamefont {Sknepnek}},\ and\ \bibinfo {author} {\bibfnamefont {E.}~\bibnamefont {Bertin}},\ }\bibfield  {title} {\bibinfo {title} {Dense active matter model of motion patterns in confluent cell monolayers},\ }\href@noop {} {\bibfield  {journal} {\bibinfo  {journal} {Nat. Commun.}\ }\textbf {\bibinfo {volume} {11}},\ \bibinfo {pages} {1405} (\bibinfo {year} {2020})}\BibitemShut {NoStop}%
\bibitem [{\citenamefont {Lifshitz}\ and\ \citenamefont {Kolsky}(1964)}]{LK1964}%
  \BibitemOpen
  \bibfield  {author} {\bibinfo {author} {\bibfnamefont {J.~M.}\ \bibnamefont {Lifshitz}}\ and\ \bibinfo {author} {\bibfnamefont {H.}~\bibnamefont {Kolsky}},\ }\bibfield  {title} {\bibinfo {title} {Some experiments on anelastic rebound},\ }\href@noop {} {\bibfield  {journal} {\bibinfo  {journal} {Journal of the Mechanics and Physics of Solids}\ }\textbf {\bibinfo {volume} {12}},\ \bibinfo {pages} {35} (\bibinfo {year} {1964})}\BibitemShut {NoStop}%
\bibitem [{\citenamefont {Sandeep}\ \emph {et~al.}(2021)\citenamefont {Sandeep}, \citenamefont {Senetakis}, \citenamefont {Cheung}, \citenamefont {Choi}, \citenamefont {Wang}, \citenamefont {Coop},\ and\ \citenamefont {Ng}}]{SSC+2021}%
  \BibitemOpen
  \bibfield  {author} {\bibinfo {author} {\bibfnamefont {C.~S.}\ \bibnamefont {Sandeep}}, \bibinfo {author} {\bibfnamefont {K.}~\bibnamefont {Senetakis}}, \bibinfo {author} {\bibfnamefont {D.}~\bibnamefont {Cheung}}, \bibinfo {author} {\bibfnamefont {C.~E.}\ \bibnamefont {Choi}}, \bibinfo {author} {\bibfnamefont {Y.}~\bibnamefont {Wang}}, \bibinfo {author} {\bibfnamefont {M.~R.}\ \bibnamefont {Coop}},\ and\ \bibinfo {author} {\bibfnamefont {C.~W.~W.}\ \bibnamefont {Ng}},\ }\bibfield  {title} {\bibinfo {title} {Experimental study on the coefficient of restitution of grain against block interfaces for natural and engineered materials},\ }\href@noop {} {\bibfield  {journal} {\bibinfo  {journal} {Canadian Geotechnical Journal}\ }\textbf {\bibinfo {volume} {58}},\ \bibinfo {pages} {35} (\bibinfo {year} {2021})}\BibitemShut {NoStop}%
\bibitem [{\citenamefont {Besold}\ \emph {et~al.}(2000)\citenamefont {Besold}, \citenamefont {Vattulainen}, \citenamefont {Karttunen},\ and\ \citenamefont {Polson}}]{BVK+2000}%
  \BibitemOpen
  \bibfield  {author} {\bibinfo {author} {\bibfnamefont {G.}~\bibnamefont {Besold}}, \bibinfo {author} {\bibfnamefont {I.}~\bibnamefont {Vattulainen}}, \bibinfo {author} {\bibfnamefont {M.}~\bibnamefont {Karttunen}},\ and\ \bibinfo {author} {\bibfnamefont {J.~M.}\ \bibnamefont {Polson}},\ }\bibfield  {title} {\bibinfo {title} {Towards better integrators for dissipative particle dynamics simulations},\ }\href@noop {} {\bibfield  {journal} {\bibinfo  {journal} {Phys. Rev. E}\ }\textbf {\bibinfo {volume} {62}},\ \bibinfo {pages} {R7611} (\bibinfo {year} {2000})}\BibitemShut {NoStop}%
\end{thebibliography}%

\end{document}